\documentclass{emulateapj}
\def\Sref#1{\S\ref{Sec:#1}}
\def\Fref#1{Figure~\ref{Fig:#1}}
\def\Tref#1{Table~\ref{Table:#1}}

\newcommand{\altm} {\altaffilmark}
\newcommand{\psr}  {J0737$-$3039}
\newcommand{\hh}   {^{\mathrm h}}  
\newcommand{\mm}   {^{\mathrm m}}  
\newcommand{\dd}   {^\circ}  
\newcommand{\am}   {^\prime}  

\shorttitle{Radio Emission from \psr\ Revisited}
\shortauthors{Chatterjee, Goss, \& Brisken}
\slugcomment{Accepted by the Astrophysical Journal Letters, 2005 October 13}

\begin{document}
\title{Radio Emission from the Double-Pulsar System J0737$-$3039 Revisited}
\author{
S. Chatterjee\altm{1}, W. M. Goss\altm{2}, and W. F. Brisken\altm{2}
}

\begin{abstract}

The double pulsar \psr\ is the only known system in which the
relativistic wind emitted by a radio pulsar demonstrably interacts
with the magnetosphere of another one.  We report radio
interferometric observations of the \psr\ system with the VLA at three
wavelengths, with each observation spanning a full binary orbit.  We
detect \psr\ at 1.6 and 4.8~GHz, derive a spectral index of $-2.3 \pm
0.2$, and place an upper limit on its flux density at 8.4~GHz.
Orbital modulation is detected in the 1.6~GHz data, with a significance
of $\sim 2\sigma$.  Both orbital phase-resolved and phase-averaged
measurements at 1.6~GHz are consistent with the entire flux density
arising from the pulsed emission of the two pulsars.  Contrary to
prior results, we find no evidence for unpulsed emission, and limit it
to less than 0.5~mJy ($5 \sigma$).

\end{abstract}


\altaffiltext{1}{Jansky Fellow, National Radio Astronomy Observatory;
        and Harvard-Smithsonian Center for Astrophysics, 60 Garden Street,
        Cambridge, MA 02138;  schatterjee@cfa.harvard.edu}
\altaffiltext{2}{National Radio Astronomy Observatory, P.O. Box O,
   Socorro, NM 87801; mgoss@aoc.nrao.edu,wbrisken@aoc.nrao.edu} 

\section{Introduction}\label{Sec:intro}

The double pulsar \psr\ is one of the most extraordinary systems in
all of astronomy.  It consists of a recycled pulsar
\citep[``A'';][]{bdp+03} with a period of 22.7~ms, in a 2.4~hr
eccentric binary orbit ($e=0.09$) with a young pulsar
\citep[``B'';][]{lbk+04} that has a 2.8~s period.  Since both neutron
stars in the highly relativistic binary system are detected as radio
pulsars, unprecedented tests of theories of gravitation are
possible. In addition, the orbit is tight enough that the relativistic
wind from A interacts with the magnetosphere of B, providing a unique
laboratory for studying the energy dissipation of neutron stars.

The \psr\ system offers an astonishing variety of observational
surprises \citep{lbk+04}.  The wind from A apparently regulates the
pulsed emission from B, which is bright only at certain parts of its
orbit. (In fact, this intermittency was responsible for B not being
detected simultaneously with A.)  Single pulses from B show features
drifting at the beat frequency between the periods of the two pulsars,
reflecting the direct impact of electromagnetic radiation from A on B
\citep{mkl+04}.  Meanwhile, since the plane of the orbit is almost
edge-on to our line of sight, A undergoes a short eclipse
\citep{lbk+04,krb+04} when it passes behind B.  In a remarkable
analysis, \citet{mll+04} showed that the eclipse of A is modulated at
half the rotational period of B and ascribed the phase-dependent
opacity to synchrotron absorption in a cometary magnetosheath (as
suggested by \citealt{as-HEAD04}) that surrounds and rotates with B.
More recently, \citet{lt05} suggest that relativistically hot dense
plasma confined by the magnetic field of pulsar B scatters emission
from A at synchrotron resonance, leading to the observed
phase-dependent opacity.

Based on an ATCA (Australia Telescope Compact Array) observation,
\citet{bdp+03} reported that the radio flux density from \psr\ was
$6.9 \pm 0.6$~mJy at 1.4~GHz. (As is standard practice, all pulsar
flux densities discussed here are averaged over pulse phase.)  With
the discovery of B, \citet{lbk+04} measured individual pulsar flux
densities of $1.6 \pm 0.3$~mJy for A and $(0 - 1.3) \pm 0.3$~mJy for
B, for an orbital phase-averaged pulsed flux density of $\sim 1.8$~mJy
at 1.4~GHz.  They suggested that the remaining $\sim 5$~mJy was
unpulsed emission that arose from the interaction of the relativistic
winds of A and B.  Such an interaction also provides a possible
explanation for the X-ray emission from the system, as detected by
{\sl Chandra} in a brief (10~ks) observation \citep{mcb+04}. Somewhat
deeper {\sl XMM-Newton} observations \citep{cpb04,pdm+04} show no
evidence for any modulation in the X-ray emission, and the existing
X-ray data do not necessarily require interaction between the two
pulsars, but deeper observations (planned for late 2005) will be
required for a definite conclusion.

Much theoretical effort has gone into interpreting the unusual
richness of behavior exhibited by \psr\ 
\citep[e.g.,][]{lyutikov04,drb+04}.  \citet{as-HEAD04} suggest that
the unpulsed radio emission from \psr\ may be synchrotron emission
from the magnetosheath of B enhanced by gyrophase bunching, while
\citet{tt04} propose that it arises either from a shock-heated
electron cloud, or from a bow shock surrounding B's cometary
magnetosheath.  Either scenario has interesting implications for our
understanding of relativistic winds from neutron stars.  We note,
however, that the existence of unpulsed emission is not required by
models \citep[e.g.,][]{lt05}.

\begin{deluxetable*}{ccccccccc}
\tablecolumns{9}
\tablewidth{0pc} 
\tablecaption{Summary of Observations\label{Table:obs}}
\tablehead{ 
\colhead{Date} & \colhead{Array} &
\colhead{$\nu$} & \colhead{$\Delta \nu$} &
\colhead{T$_{\rm int}$} & \colhead{Beam} & \colhead{P.A.} &
\colhead{$S_\nu$} & \colhead{$\sigma_\nu$} \\
\colhead{} & \colhead{} & \colhead{(MHz)} & \colhead{(MHz)} &
\colhead{(hr)} &
\colhead{($\arcsec$)} & \colhead{($\arcdeg$)} &
\colhead{(mJy)} & \colhead{(mJy)}
}
\startdata 
2003-Jun-28 & ATCA & 1384 & 104 & 4.8 & $10.6 \times 4.7$  & $-12.2$ & 1.83 & 0.34 \\
2004-Aug-01 & VLA-B & 1465 & 20 & 0.7 & $12.0 \times 3.5$  & $-0.7 $ & 2.28 & 0.18 \\
2005-Nov-28 & VLA-A & 1666 & 20 & 2.2 & $2.25 \times 0.82$ & $ 4.0 $ & 1.88 & 0.10 \\
2005-Nov-29 & VLA-A & 4885 & 44 & 2.2 & $1.33 \times 0.37$ & $-10.4$ & 0.17 & 0.03 \\
2005-Nov-30 & VLA-A & 8435 & 44 & 2.0 & $0.33 \times 0.26$ & $ 0.2 $ & ---  & 0.02 
\enddata 

\tablecomments{T$_{\rm int}$ is the total on-source integration time, while
  the span of the data is longer. P.A. is the beam position angle,
  measured east of north.}

\end{deluxetable*} 

In this {Letter} we report radio interferometric observations of
the \psr\ system with the NRAO {Very Large Array} (VLA) at three
frequency bands, with each observation spanning a full pulsar orbit.
Interferometric observations are sensitive only to the total system
flux density, which we determine at 1.6 and 4.8~GHz and limit at
8.4~GHz.  We derive a spectral index ($S_\nu \propto \nu^{\alpha}, \:
\alpha = -2.3 \pm 0.2$), which is consistent with radio emission from
pulsars, but steeper than that expected for a pulsar wind nebula.
There is no evidence for extended structure, as might be expected for
a pulsar wind nebula.  At 1.6~GHz, we detect modulation due to the
orbital phase-dependent emission from B (with a significance of $\sim
2 \sigma$) and find upper limits on the system flux density when A is
in eclipse.  Crucially, the entire flux density at 1.6~GHz can be
accounted for by the sum of pulsed flux densities from A and B, as
reported by \citet{lbk+04}.  Thus, there is {\em no evidence for
unpulsed radio emission} from the system.  Each aspect is discussed in
detail below.

\section{Flux Density and Spectral Index}\label{Sec:sp}

\psr\ was observed with the VLA in its widest configuration (A array)
on 2004 November 28, 29, and 30, at 1.6, 4.8, and 8.4~GHz
respectively, with observations spanning a full binary orbit on each
day. The observational parameters are summarized in \Tref{obs}. In
each case, 3C147 was observed as the primary flux density calibrator,
and a nearby compact extragalactic source was employed as a phase
calibrator and secondary amplitude calibrator.  At 1.6~GHz, we
employed J0738$-$3322 (2\fdg7 away), while J0747$-$3310 (3\fdg2 away)
was used at 4.8 and 8.4~GHz.

\begin{deluxetable*}{ccrrl}
\tablecolumns{5}
\tablewidth{0pc} 
\tablecaption{Sources in the Field of \psr\ at 1.6~GHz\label{Table:src}}
\tablehead{ 
\colhead{RA} & \colhead{Dec} & 
\colhead{Flux Density} & \colhead{$\Delta\theta$} &
\colhead{Comment} \\
\colhead{(J2000)} & \colhead{(J2000)} & 
\colhead{(mJy/beam)} & \colhead{(\arcmin)} &
\colhead{}
}
\startdata 
$07\hh 37\mm 51\fs248$ & $-30\dd 39\am 40\farcs83$ &    1.9 &  0.00 & Pulsar \\
$07\hh 37\mm 51\fs075$ & $-30\dd 44\am 03\farcs55$ &    2.0 &  4.38 & Compact double \\
$07\hh 37\mm 48\fs034$ & $-30\dd 33\am 33\farcs90$ &    0.8 &  6.15 &  \\
$07\hh 37\mm 52\fs411$ & $-30\dd 47\am 11\farcs04$ &    1.2 &  7.51 &  \\
$07\hh 37\mm 44\fs819$ & $-30\dd 48\am 55\farcs55$ &    0.5 &  9.35 &  \\
$07\hh 38\mm 30\fs425$ & $-30\dd 46\am 49\farcs25$ &    2.2 & 11.04 &  \\
$07\hh 38\mm 43\fs344$ & $-30\dd 40\am 40\farcs16$ &    1.4 & 11.25 &  \\
$07\hh 38\mm 40\fs997$ & $-30\dd 44\am 45\farcs52$ &    5.6 & 11.84 &  \\
$07\hh 38\mm 46\fs237$ & $-30\dd 36\am 50\farcs10$ &    2.3 & 12.17 & Extended double \\
$07\hh 37\mm 42\fs214$ & $-30\dd 27\am 19\farcs34$ &    3.3 & 12.51 &  \\
$07\hh 36\mm 51\fs230$ & $-30\dd 37\am 46\farcs16$ &    3.3 & 13.05 &  \\
$07\hh 37\mm 44\fs153$ & $-30\dd 54\am 36\farcs49$ &    1.3 & 15.00 &  \\
$07\hh 37\mm 09\fs591$ & $-30\dd 27\am 10\farcs19$ &    146 & 15.39 & Partly resolved \\
$07\hh 38\mm 19\fs790$ & $-30\dd 25\am 05\farcs23$ &   1226 & 15.83 & Partly resolved \\
$07\hh 36\mm 45\fs369$ & $-30\dd 27\am 18\farcs78$ &     18 & 18.82 &  \\
\enddata 

\tablecomments{Flux densities are from VLA A-array observations at
  1.6~GHz, and an epoch of MJD~53,337, corrected for the effects of the
  primary beam. $\Delta\theta$ is the angular offset in arcminutes from
  the pulsar position.  Absolute positions are accurate to $\sim 0\farcs05$.}

\end{deluxetable*} 

The data were processed within {AIPS}, the Astronomical Image
Processing System.  The steps included data flagging to excise
radio-frequency interference, setting the flux density scale based on
observations of 3C147, and determining antenna gains and phases from
scans on the secondary calibrator.  At 1.6~GHz, two passes of
self-calibration were employed, with all detected sources within the
primary beam included in the model.  The positions and flux densities
of all strong sources in the field are listed in \Tref{src}.

\psr\ was detected at 1.6 and 4.8~GHz, and we fit elliptical Gaussians
to the images (task {\tt JMFIT} in AIPS) to determine the source
position and flux density. The source flux density is $1.88 \pm
0.10$~mJy at 1.6~GHz and $0.17 \pm 0.03$~mJy at 4.8~GHz, which leads
to a spectrum $S_\nu \propto \nu^{\alpha}, \; \alpha = -2.3 \pm 0.2$,
as plotted in \Fref{sp}.  At 8.4~GHz, we do not detect the pulsar, but
place an upper limit of 0.06~mJy ($3\sigma$) on its flux density.  As
shown in \Fref{sp}, such an upper limit is consistent with the
measured spectral index.

\begin{figure}[ht]
\plotone{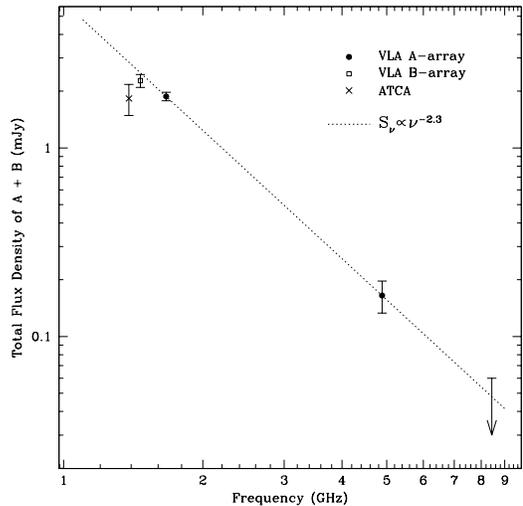}
\caption{Spectrum of \psr. With VLA A-array observations spanning
a full binary orbit at each frequency, we determine source flux
densities of $1.88 \pm 0.10$~mJy at 1.6~GHz and $0.17 \pm 0.03$~mJy at
4.8~GHz, and place a $3\sigma$ upper limit of 0.06~mJy at 8.4~GHz. A
spectral fit is shown based on the two detections.  In addition,
detections of \psr\ in archival data from VLA B-array and ATCA
observations at 1.4~GHz are also plotted.  The flux density
determinations from archival data are consistent with the derived
spectrum (see discussion in \Sref{discuss}).  
}
\label{Fig:sp}
\vspace{-5mm}
\end{figure}

 
\section{Orbital Modulation of the Radio Emission}\label{Sec:orbit}

The shortest integration time possible at the VLA is 1.66~s, so that
the rotational phase of either pulsar A or B is impossible to resolve.
However, the 1.6~GHz VLA observations have sufficient signal-to-noise
ratio (S/N) to permit estimates of the system flux density over
several bins in binary orbital phase.  Based on the observations
presented by \citet{bpm+05}, B is expected to be detectably brighter
(i.e., to cause the \psr\ system to have a detectably higher flux
density) between orbital phases of $\sim$274 and 297\arcdeg, and
brightest between $\sim$196 and 224\arcdeg, at the time of our VLA
observations (2004 November).  Meanwhile, A undergoes its brief
eclipse at an orbital phase of 90\arcdeg.  (In each case, orbital
phase is measured as the sum of the longitude of periastron and the
true anomaly of the respective pulsars, so that A and B are always
180\arcdeg\ apart in their orbital phase.)

In order to estimate the flux density of \psr\ over bins in orbital
phase, all the field sources first have to be accounted for.  An
interferometer samples different points in the visibility (spatial
frequency) plane as the Earth rotates over the course of an
observation, leading to a time-varying beam pattern.  We account for
this effect by first imaging the entire data span (as described above)
and then subtracting the modeled contributions of each background
source from the interferometric visibilities.  The remainder consists
of the contributions of the two pulsars, A and B, along with the
thermal noise and any systematic errors due to unmodeled source
contributions\footnote{Such errors may arise, for example, due to
modulation of strong sources near the edge of the primary beam because
of pointing errors.}.  These residuals were binned in time, using
published pulse timing solutions \citep{lbk+04} to predict the orbital
phase as observed from the VLA.  Finally, images were created from the
data in each time bin, and an elliptical Gaussian at a known (fixed)
position was fit to the image to determine the flux density of \psr.

\begin{figure*}[th]
\epsscale{1.1}
\plottwo{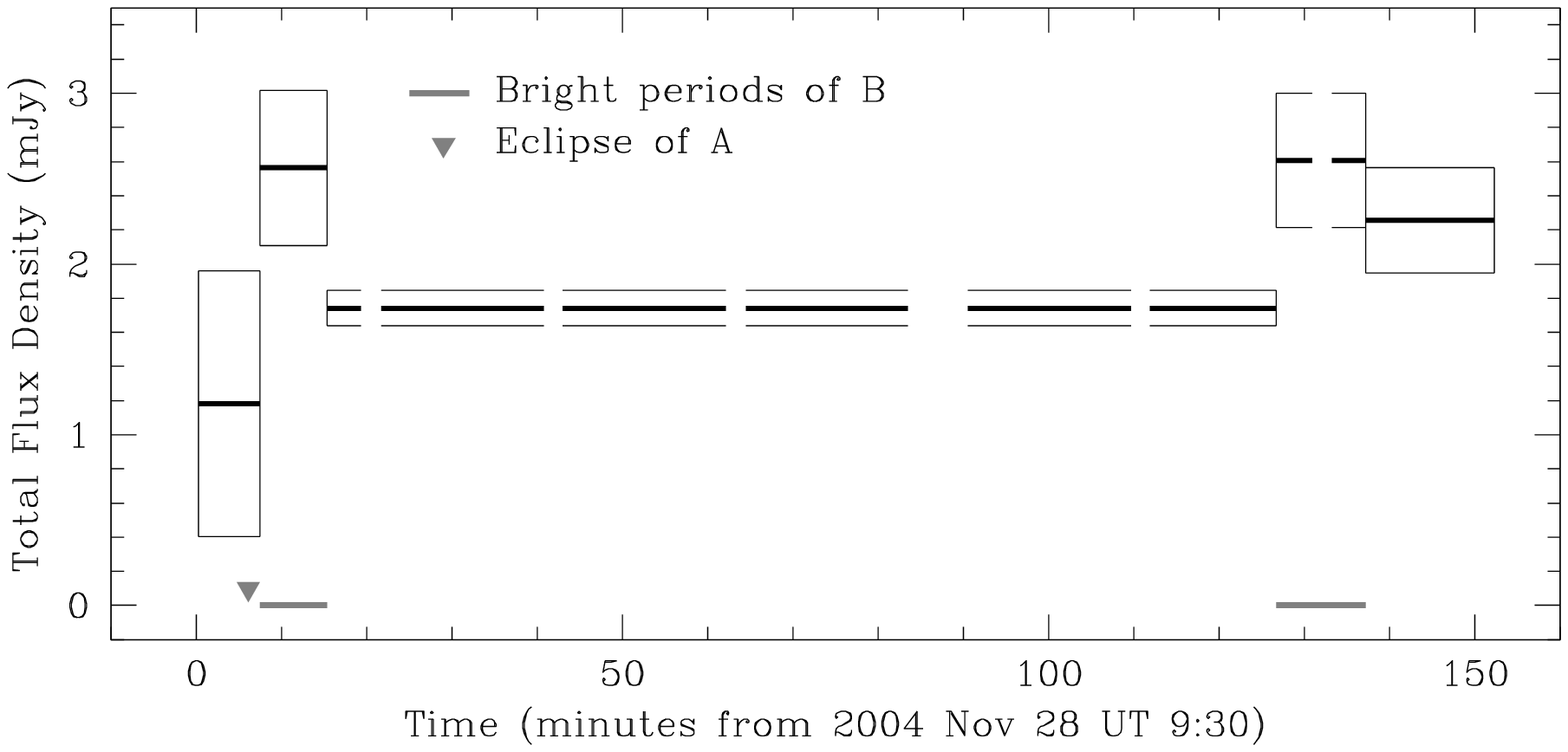}{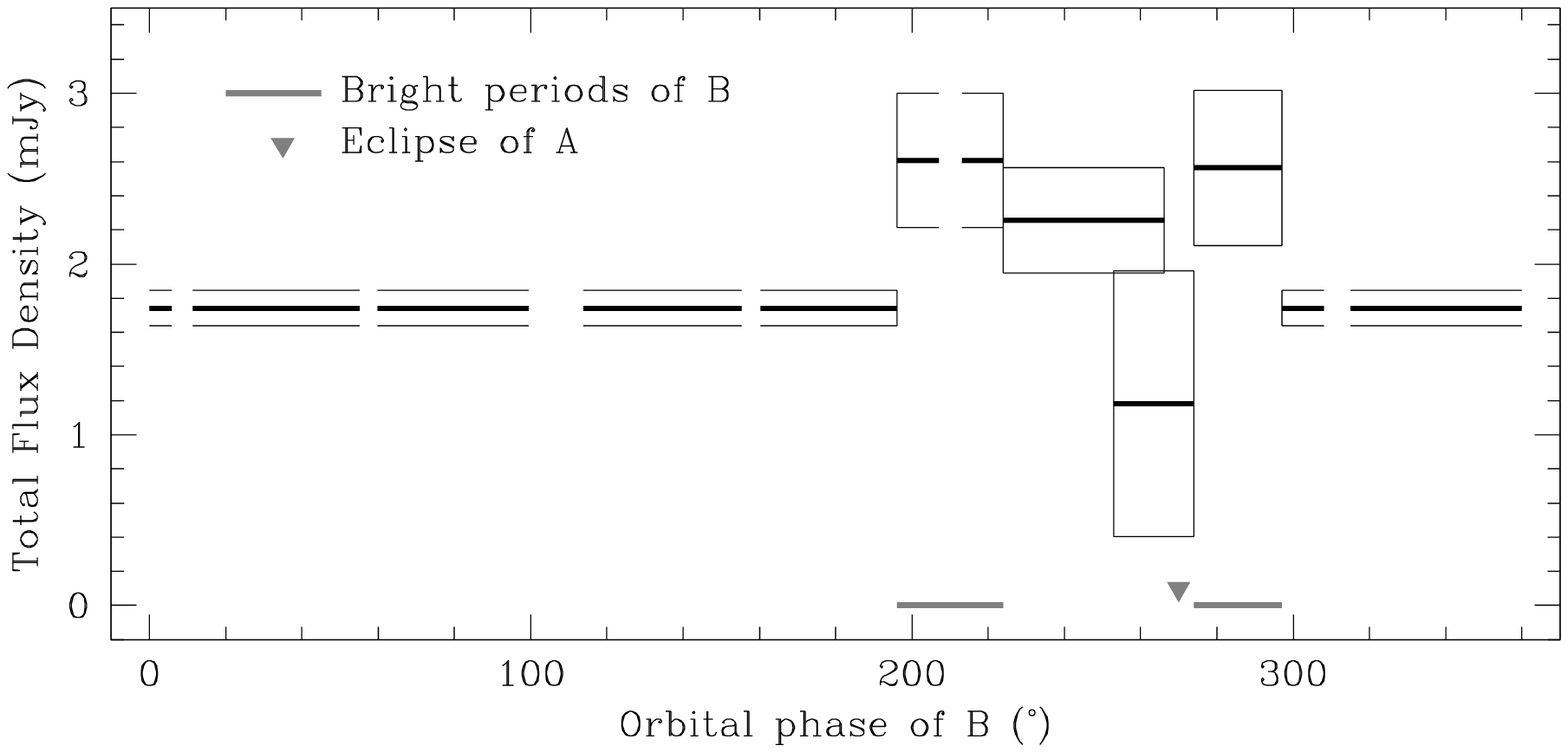}
\caption{ Light curve of \psr\ at 1.6~GHz, as a function of time
(left) and orbital phase of B (right). The flux density is
indicated (thick line) along with $\pm 1\sigma$ errors on the
measurement (thin lines). The observations span slightly more than a
full orbit, with gaps corresponding to interleaved calibration scans.
The times and phases when B is expected to be brighter are identified.
The flux density varies from $2.6\pm0.4$ to $1.7\pm0.1$~mJy, due to
the expected variability of pulsar B, which we thus detect at the
$\sim 2\sigma$ level. \psr\ is not detected during the $\sim 1$~minute
eclipse of A (when A is at an orbital phase of 90\arcdeg, and B is at
a phase of 270\arcdeg, as marked), but we lack the S/N to derive
significant upper limits on the flux density in that interval.}
\label{Fig:lc}
\end{figure*}

Orbital phase-resolved flux densities at 1.6~GHz are plotted in
\Fref{lc}.  We find that the flux density of the \psr\ system (A + B)
is $2.6\pm0.4$~mJy when B is brightest (orbital phase
$\sim$196--224\arcdeg), $2.6\pm0.5$~mJy when it is weakly detectable,
and $1.7\pm0.1$~mJy when it is faint and only A is expected to be
visible.  These estimates are completely consistent with the
measurements of \citet{lbk+04}, as listed in \Sref{intro}.  We also
created short time bins at and around the eclipse of A.  While \psr\
was not detected in the bin corresponding to the eclipse of A, we lack
the S/N to place a significant upper limit on the total flux
density in that short ($\sim 1$ minute) time interval.

\section{Discussion: Is there Unpulsed Radio Emission?}\label{Sec:discuss}

At 1.6~GHz, we find an orbital phase-averaged flux density of
$1.88\pm0.10$~mJy for \psr, consistent with the sum of the pulsed
emission from A and (intermittently) B.  Our phase-resolved
measurements are also consistent with this picture.  After subtracting
the pulsed flux density \citep[$\sim$1.8~mJy;][]{lbk+04}, there is no
evidence for any unpulsed flux, and we can rule out the existence of
unpulsed emission at the $0.5$~mJy level ($5\sigma$).

Given the discrepancy with previous work, we have examined the
archival ATCA data that form the basis for the claimed unpulsed
emission, as well as archival VLA B-array data at 1.4~GHz (see
\Tref{obs}). The VLA B-array data span only 40~minutes (when B spans
an orbital phase range of 96\arcdeg\ to 197\arcdeg), and after
standard calibration, we find a flux density of $2.3\pm0.2$~mJy for
\psr.  As shown in \Fref{sp}, this is consistent with the spectrum
derived from 1.6 and 4.8~GHz VLA A-array observations.  There is also
no evidence for any extended structure, and the mean flux density is
$< 30 \mu$Jy~beam$^{-1}$ in an annulus between 10\arcsec\ and
100\arcsec\ from the pulsar.

The ATCA data were calibrated within MIRIAD, and then imaged in AIPS.
Due to limited coverage of the visibility plane, the data are
comparatively more difficult to self-calibrate and image.  After
discarding data in the inner visibility plane and only retaining data
between 5 and 20~k$\lambda$, we find a flux density of $1.8\pm0.3$~mJy
for \psr, inconsistent with the value of $6.9\pm0.6$~mJy reported by
\citet{bdp+03}.  We suggest three possible causes for the discrepancy:
(1) there are many sources in the field (\Tref{src}), and it is
possible, though unlikely, that the flux density may have been
transcribed for an incorrect one; (2) the ATCA data on \psr\ were
acquired in pulsar binning mode, while data on the calibrators
(B1934$-$638 and B0727$-$365) were acquired in the standard mode,
necessitating the scaling of amplitudes to obtain correct flux
densities. We have averaged all the pulsar bins in our analysis of the
ATCA data to avoid the need for scaling. And (3) there may be a very
faint extended nebula surrounding the pulsar, and contributing flux
density on scales $\gtrsim 40\arcsec$, which we have excised but that
\citet{bdp+03} retained. The last scenario is rather unlikely, but
cannot be definitively ruled out yet.  In any case, the VLA
observations, acquired at different times, frequencies, and array
configurations, provide a more reliable estimate of the flux density
of \psr\ itself.

We have thus detected \psr\ at 1.6 and 4.8~GHz, derived a spectral
index of $-2.3 \pm 0.2$, and placed an upper limit on its flux density
at 8.4~GHz. Orbital modulation due to B is detected in the 1.6~GHz
data.  Both orbital phase-resolved and phase-averaged measurements at
1.6~GHz are consistent with the entire flux density arising from 
the pulsed emission of A and B.  We find no evidence for unpulsed
emission, and limit it to less than 0.5~mJy ($5 \sigma$).  This is a
disappointing conclusion for theories that attempted to relate the
unpulsed radio emission and X-ray emission to a common origin in a
shock created by the interaction between A and B.  If the X-ray
emission also turns out to have an ``ordinary'' origin in the
magnetosphere of A, rather than being produced by the interaction
between two relativistic winds, then the processes involved in the
energy dissipation of neutron stars will have proven more elusive than
expected.

\acknowledgements 

We thank Maura McLaughlin, Michael Kramer, and Ingrid Stairs for
helping predict the orbital phase of \psr, Joseph Gelfand for
assistance with archival ATCA data, Anatoly Spitkovsky and Maxim
Lyutikov for helpful discussions, and the anonymous referee for 
helpful comments.  S. C. is a Jansky Fellow of the National Radio
Astronomy Observatory (NRAO).  NRAO is a facility of the National
Science Foundation, operated under cooperative agreement by Associated
Universities, Inc.


\end{document}